Effects of Misorientation on Single Crystal Plasticity by Finite Element Methods

John D. Shimanek[1,*], Zi-Kui Liu[1], Allison M. Beese[1,2,*]

[1] Department of Materials Science and Engineering, The Pennsylvania State University, University Park, PA, 16802, USA

[2] Department of Mechanical Engineering, The Pennsylvania State University, University Park, PA 16802, USA

* Corresponding authors: jshimanek@psu.edu, amb961@psu.edu

**Abstract**:

The crystal plasticity finite element method (CPFEM) is a widely used technique for resolving macroscopic stress and strain onto the physically relevant length scales of grains and slip systems in ductile crystalline materials like structural metals. Here, the calibration of a CPFEM model for single crystal applications was found to depend critically on loading orientation, with an effect significant even at an angle of 0.1°. Slight misorientation from high symmetry loading affected lattice rotation during tensile deformation, changing the number of active slip systems, and, as a result, the overall stress-strain behavior. The strongest misorientation effects occurred around the multi-slip orientations of [001], [111], [101], and [102], while the single slip orientation of [213] showed a negligible effect, as expected, and the double slip orientation of [112] showed less of an effect than [102] due to its relative lattice orientation stability. The magnitude of the misorientation effect increased dramatically with the strength of slip system interaction, which, in the chosen hardening framework, is represented by the latent hardening coefficient. In a case study on [001] Cu, offsets of 0.3–2.0° gave stress values at an engineering strain of 0.25 that were lower by 15–18% relative to the direct loading values, highlighting the importance of exact orientations for single crystal plasticity parameterization and application.

## 1. Introduction

Crystal plasticity finite element (CPFE) methods relate macroscopic states of stress and strain to a fundamental understanding of microscopic deformation mechanisms, especially the glide of dislocations within slip planes [1]. These models are often parameterized by fitting to the stress-strain response of the bulk material of interest, either in single crystal or polycrystal form [2]. While the physical underpinning of CPFE models lends some plausibility to their predictive power in extrapolative situations, careful parameterization is especially critical for such applications. In the case of single crystal metallic materials, data from uniaxial deformation experiments in one or more crystallographic orientations are often used to fit the material-specific parameters of the CPFE model [3], although the transferability of parameters fit on only one orientation can be poor [4]. These experiments inherently include some slight yet unknown deviation from the nominal loading orientation, but computational models do not. As a result, the direct comparison between experimental and computational results for the purposes of determining material properties is fraught. Here, we examine the effect of slightly offset loading orientations on the behavior of single crystal CPFE models and the resulting misestimation of material behavior, focusing especially on the substantial impact of slip system interactions.

Early experimental studies noted the general importance of small offsets in the overall tensile behavior of single crystal specimens, ascribing softer stress-strain behavior to small offsets from high-symmetry tensile orientations [5], and such effects are expected to be correctly represented in the computational modeling of crystal plasticity [6]. Initial strain hardening models were phenomenological, their forms chosen to match experimental results [7], and their treatment of slip system interactions embodied in a single latent hardening ratio of between 1.0 and 1.4 [8]. These models were later joined by more elaborate hardening frameworks using dislocation densities as a set of internal variables to describe the strength of each slip system [9]. In most dislocation density based hardening laws, slip system interaction effects appear in both the coupled density evolution equations and in the relation between densities and slip system strength, complicating a direct comparison to the constant value used in phenomenological laws. To reduce the misorientation effect on the parameterization of these models to high-symmetry loading orientation data, they can be parameterized with a small offset [10]. An exploration of slip system activity for small offsets has been performed by Cuitiño and Ortiz, who also discussed the latent hardening effects of their dislocation density based model but left out the connection between misorientation and latent hardening modeling [9]. Slip system interaction strengths have been discussed in the contexts of both crystal plasticity and dislocation dynamics [11–13], although the connection to model parameterization and misorientations has not been explicitly made.

The acknowledged importance of misorientation effects, along with the continued use of phenomenological hardening models in CPFE modeling, warrants a systematic examination of the effect and especially of its interplay with the latent hardening ratio. The current practices in CPFE applications largely depend on single orientation data with no assumed offset for the parameterization of single crystal hardening models [14], or, more commonly, parameterized to the specialized geometry of an experimental design of interest [15]. In the case of phenomenological models, the choice of latent hardening ratio is often justified only through passing reference to prior use in the literature. These choices can have considerable impacts on the resulting material parameters depending on the specific CPFE framework.

While the misorientation effect is expressly important to CPFE methods, modern studies using molecular dynamics run into a similar phenomenon of offsets influencing the competition between slip systems during deformation, especially during uniaxial loading when the dominant slip system influences the lattice rotation that then determines the resolved shear stress on other slip systems [16]. Further corroboration of the importance of slip system interaction and its influence on lattice rotation has been demonstrated with large scale molecular dynamics of several crystal orientations [17]. Examples are also found in modern experiments, with slight misorientations in the modeling of Taylor impact samples being shown to better match experimental results [18]. However, the most related area of research remains CPFE modeling, especially the development of more robust and predictive frameworks. In this domain, the misorientation effect mainly acts to increase the convolution of geometry and material, which should be minimized to allow for more predictable material parameters that may eventually be useful in an alloy design framework.

The following sections present a systematic examination of misorientation effects within phenomenological CPFE models of pure Cu single crystals and their strong dependence on latent hardening. First, the chosen CPFE methodology and hardening model are outlined along with several deformation metrics whose evolution with respect to loading orientation offset angle inform the physical interpretation of the effect. The results of uniaxial tension calculations along multiple loading orientations are then discussed through each deformation metric. Finally, the severity of the misorientation effect in single crystal plasticity modeling is addressed in its close relation to the latent hardening ratio.

2. **Methods**

The crystal plasticity material model is implemented through an existing user-defined material subroutine to the commercially available Abaqus finite element software. The subroutine is that of Huang [19], and it follows an established constitutive framework for crystal plasticity modeling [6], which will be briefly outlined here. The overall deformation gradient, $\boldsymbol{F}$, is first decomposed multiplicatively into elastic and plastic components as

$$\boldsymbol{F} = \boldsymbol{F}^* \cdot \boldsymbol{F}^P \tag{1}$$

where $F^*$ is the elastic component of the deformation gradient, and only the plastic portion, $\boldsymbol{F}^P$, is associated with shearing on slip systems. Thus, the contribution from plasticity can be related to its rate of change, $\dot{\boldsymbol{F}}^P$, and the slipping rate on each slip system as

$$\dot{\boldsymbol{F}}^P \cdot (\boldsymbol{F}^P)^{-1} = \sum_{\alpha=1}^{N} \dot{\gamma}^{(\alpha)} \boldsymbol{s}^{(\alpha)} \boldsymbol{m}^{(\alpha)} \tag{2}$$

where $\dot{\gamma}^{(\alpha)}$ is the slipping rate, $\boldsymbol{s}^{(\alpha)}$ the slip direction, and $\boldsymbol{m}^{(\alpha)}$ the slip plane, all on system $\alpha$ of a total $N$ slip systems. Since the present study focuses on a face-centered cubic material, $N = 12$.

The slipping rate on each system is given as

$$\dot{\gamma}^{(\alpha)} = \dot{\gamma}_0 \left| \frac{\tau^{(\alpha)}}{\tau_c^{(\alpha)}} \right|^{\frac{1}{m}} \text{sgn}(\tau^{(\alpha)}) \tag{3}$$

where $\dot{\gamma}_0$ is a reference strain rate associated with the strain rate exponent $m$, and $\tau^{(\alpha)}$ is a resolved shear stress on the slip system with a critical value of $\tau_c^{(\alpha)}$. The critical value represents slip system strength and evolves with shear strain according to the material hardening law.

Although there are many potential hardening laws capable of capturing experimental deformation data, the form that is chosen for the present calculations is that from Pierce, Asaro, and Needleman (PAN) [7]:

$$\dot{\tau}_c^{(\alpha)} = \sum_{\beta=1}^{N} h_{\alpha\beta} |\dot{\gamma}^{(\beta)}| \tag{4}$$

where the form of the hardening matrix is

$$h_{\alpha\beta} = \left(\delta_{\alpha\beta} + q(1 - \delta_{\alpha\beta})\right) h_0 \operatorname{sech}^2 \left|\frac{h_0 \gamma}{\tau_s - \tau_0}\right| \tag{5}$$

Here, $q$ is the latent hardening ratio ($q$-value) representing the hardening effect from slip on one system on the strength of other systems relative to its effect on the strength of the system on which slip is occurring. The model also relies on material parameters $\tau_0$, representing the initial strength of the slip systems; $h_0$, the initial hardening rate; and $\tau_s$, the saturation strength of the slip system. Calculations performed here used parameters calibrated to tensile data from single crystal Cu along the [001] loading orientation, as shown in Table 1 [20]. Additional calibrations to tensile data from multiple orientations is discussed in the supplemental material.

Although the results presented here were generated using this specific choice of hardening model, the effect is expected to be present in other hardening laws as well since they all aim to capture the same slip system level behavior. Important for attributing physical meaning to model changes within a neighborhood of parameter space, a relatively simple hardening model ameliorates the issue of non-unique parameterization, where multiple regions of parameter space perform equally well in their ability to reproduce experimental stress-strain curves. More complex hardening models, while able to capture the same stress-strain behavior, would require a more involved procedure to support proper parameterization. For instance, in the dislocation density based hardening model of Ref. [10], slip system interaction terms appear not only in their effect on the strength evolution of each slip system but also in the evolution of the dislocation density of each system, which also affects its strength. Other models account for the relative strengthening effects of various binary slip system interactions, as characterized by dislocation dynamics simulations [13,21]. These models would also benefit from consideration of the effect that slight misorientations have on their parameterization, the magnitude of which may depend on their accounting of slip system interactions. The present model treats each slip system interaction equally, allowing for simple adjustment of overall slip system interaction strength.

Models of single crystal plasticity need to account for the lattice rotation that is expected to arise from the fixed lateral motion of the grips during a uniaxial tension test. Although this can be accomplished by a realistic model of the experimental test geometry, it is also allowed in simplified models with boundary conditions that allow for tilting of the horizontal faces normal to the tensile axis. The effect of moving from a 20,000-element finite element mesh representation of a single crystal Ni wire to a 100-element chain to a single element with appropriate boundary conditions is negligible in terms of the overall stress-strain response using the same material parameters (a difference of 1.4% in the engineering stress at an engineering strain of 70%) [22]. To enable the efficient systematic study of misorientation effects, this study used a single element model that allows for the top and bottom faces to tilt by constraining only the average of the vertical (along loading) displacement of their nodes. Rigid body motion and rotation was disallowed by fixing the pair of nodes making up one vertical edge of the model in both lateral directions and an adjacent pair in the lateral direction orthogonal to the plane connecting these two edges, as illustrated by the schematic in Figure 1. The same wire geometry model previously discussed, combined with material parameters for Cu used in the rest of this study, showed a maximum stress difference below 2% compared to the single element model similarly parameterized over a strain of 25%. Additionally, based on the lattice orientation extracted from the central element of the larger model, both models exhibit very similar lattice rotation magnitudes with respect to small loading orientation offsets. A comparison of both stress response and reorientation between the models can be found in the supplemental material.

Uniaxial tension to an engineering strain of 25% was applied as a displacement for each of the below six crystallographic orientations, which were chosen to sample the main symmetries of the FCC structure:

- [001] (8 active systems),
- [111] (6 systems),
- [101] (4 systems),
- [112] (2 systems, lies between [001] and [111]),
- [102] (2 systems, lies between [001] and [101]), and
- [213] (1 system).

Each of the orientations was also run with several small offsets ranging in increments of 0.1° up to a total tilt of 2° toward the single slip [213] direction (the exception being [213] loading itself, which was offset towards [001]). Tilting each of the input orientations towards a common direction of [213] enabled a more comparable offset degree metric between initial orientations and ensured that each orientation was being gradually changed to a single slip symmetry. A misorientation towards a single slip region is also a more representative case, being more probable than a tilt occurring exactly along the double slip boundary. The calculations were run using the PAN hardening model [7], although the effect is expected to be insensitive to the exact form of the hardening law since they capture similar behavior. A more general aspect of the material model is the latent hardening coefficient, the ratio of the strain hardening effect on other slip systems compared to that on the system generating the strain. In the literature, the latent hardening ratio is generally between 1.0 and 1.4, although a lower value is occasionally used [23]. Here, each calculation was run using $q$-values of 0.8, 1.0, and 1.4.

After running the calculations, several deformation metrics were extracted, including the stress-strain response, the slip system activity, and the magnitude of the lattice rotation. For comparison between loading orientation offset amounts, a few key quantities are defined to describe the material behavior over the entire deformation, allowing a single metric to represent each offset calculation. Taking a point-by-point relative difference in the stress-strain curves between each offset orientation and its directly loaded variant provides a stress difference scalar for each offset. Similarly, taking the mean lattice rotation over the course of the deformation gives a representative value of the lattice rotation during deformation, averaging out the detailed evolution of rotation evolution with respect to strain into a single number. Lastly, the metric $\chi$ is defined here to describe the number of active slip systems at any given macroscopic strain $\gamma$:

$$\chi(\gamma) = \sum_{\alpha} \frac{\gamma^{(\alpha)}(\gamma)}{\gamma_{\max}^{(\beta)}(\gamma)} \tag{6}$$

The strain accomplished by each slip system, $\gamma^{(\alpha)}$, is normalized by the maximum strain on any slip system, $\gamma_{\max}^{(\beta)}$, before being summed. The metric of interest for each offset is the value of $\chi$ averaged over the deformation:

$$\langle \chi \rangle = \frac{1}{N} \sum_i^N \chi(\gamma_i) \tag{7}$$

where $N$ is the total number of discrete points in strain at which the metric is evaluated, here chosen to be 100.

## 3. Results and Discussion

### 3.1. Lattice Rotation

The differences in the lattice rotations averaged over the strain range of each of the CPFE tensile calculations are shown in Figure 2. There are three basic types of behavior seen in terms of the average lattice rotation with respect to initial offset angle: a sharp jump over the first 0.1–0.2°, a flat response, and the lone sigmoidal curve of the [111] loading when $q = 1.0$. Notably, all nominal loadings except for [112] and [213] show a sharp initial jump when the latent hardening ratio is 1.4. In particular, the [001] loading changes dramatically when the latent hardening ratio is increased above unity, going from a flat response to one that increases sharply over the first 0.2°. Similarly, the increase in the lattice rotation of [102] and [111] loadings sharpen with the higher $q$-value.

While the highest symmetry loadings of [111] and [001] show a change in shape for different $q$-values, the other orientations mainly show a change in magnitude and a sharpening of the misorientation effect at small offset degrees. For instance, the next highest symmetry loading axis of [101] shows a change from gradual to immediate lattice reorientation changes as $q$-value goes from low to high. At the lowest $q$-value of 0.8, only [101] loading is strongly affected by slight misorientations, with [111] showing a slight and gradual effect and all other orientations showing similar average lattice rotations throughout deformation despite the orientation offsets. Lastly, note that both double slip loading orientations, [102] and [112], differ from each other in their rotation response to loading misorientation only at higher $q$-values; [112] is much less sensitive whereas [102] destabilizes as the $q$-value increases.

*3.2. Slip System Activity*

Since the latent hardening ratio determines the strengthening occurring on a slip system as a result of strain on all other systems except itself, the relation between $q$-value and lattice rotation may be explained through an examination of the changes in slip system activity. The strain-averaged mean number of active systems, as defined in Equations 6 and 7, is shown in Figure 3. As expected, all loadings show a drop in the effective number of active slip systems as the loading orientation is tilted away from the higher-symmetry orientations, in which many slip systems are activated equally, toward the single-slip region. The drop in effective active slip systems is highest for those orientations, e.g., [001] and [111], that activate the most slip systems during direct loading. Additionally, the higher $q$-value favors a lower number of active slip systems, often showing steep drops as the offset is increased by just 0.1˚. In contrast, the lower values of latent hardening ratios produce a slower drop in the effective number of active slip systems. Note that all orientations except for [112] and [213] show the same initial $\langle \chi \rangle$ values across different $q$-values due to the stability of their lattice orientation during deformation, which can be seen from the zero points in Figure 2 for those orientations with each $q$-value.

*3.3. Stress-Strain Response*

The culmination of the lattice rotation and slip system activation changes are reflected in the difference between the stress-strain responses of directly loaded and offset calculations, which are given in Figure 4. For the case of a $q$-value of unity, only [101] and [111] show a large difference between the slightly offset stress-strain curves, with the [111] case only becoming significant beyond misorientations of a half degree. For the more frequently used higher $q$-value, this effect is amplified so that the only orientations not showing significant differences in stress response are those of [112] and [213], which have lower symmetries than most other orientations. Notably, both the double slip orientations of [102] and [112] show different misorientation sensitivities. This is likely due to the relative instability of lattice reorientations during tensile loading, with [112] being the only orientation besides [213] to show a constant but non-zero mean reorientation magnitude regardless of latent hardening ratio.

At $q = 0.8$, the only stress-strain effect from misorientation is for the [101] orientation. A slight effect is also visible for the single slip [213] orientation, but this is due not to slip system interaction effects but to the geometrical effects of axial stress resolving onto different orientations of slip planes as the applied misorientation rotates the crystal toward a "harder" loading orientation. This effect is largely washed out in other cases by the dominating effect of the slip system interactions but is still visible for all $q$-values shown in Figure 4.

While the stress differences within each orientation curve for the case of $q = 0.8$ appear noisier relative to the cases with higher $q$-values, this is predominantly due to the lack of an offset effect that otherwise washes out the scatter in the data. The noise is likely produced during the numerical solving of the hardening equations, where errors in the non-linear yielding portion of the material response are amplified during their propagation through the rest of the loading before being converted into the single, noisy error metric shown in Figure 4.

Lastly, note that the use of relative stress differences with respect to the directly loaded case allows for the comparison of response across $q$-values. With a decreasing $q$ value and otherwise constant hardening parameters, the stress response is expected to decrease. The differences between the offset and nominal loading cases would similarly decrease if they were accounted for in absolute terms. Instead, the stress metrics presented in Figure 4 use relative stress differences from the nominal case across all values of $q$, enabling a better isolation of misorientation effects.

### 3.4. Context and Implications

Assuming a misorientation of just one or two tenths of a degree results in hardening parameters that significantly differ from the nominal case for high-symmetry loadings. This is visualized for the example of [001] loading in Figure 5, which expands upon a few of the points that make up one curve from Figure 4. The expected error of misestimating the hardening parameters is seen in the difference between the nominal loading curve and any of the slightly offset loading curves. For a model parameterized to single crystal Cu [001] tensile data without any offset, and taking the offset case to be more realistic, the stress error at 25% engineering strain is around 20 MPa above the offset value, which converges after just 0.2° offset to be around 120 MPa. The stress difference at 0.3° of offset is 15%, which is near to its final value of 18% at 2°. Larger errors are expected for higher strain levels or parameterization to [101] data while lower errors are expected for parameterizations to other orientations and for latent hardening ratios less than unity.

The potential for large differences in stress-strain responses between slight offset magnitudes indicates that parameterization of crystal plasticity models to high-symmetry tensile data requires the application of an imperfect loading orientation in the computational models. The present results suggest that, depending on the hardening model's consideration of latent hardening effects, the misorientation need only be fractions of a degree to capture the indirect loading behavior that is expected to arise naturally in experiments. Although the present results were generated using the PAN hardening model, all hardening models somehow account for slip system interactions, and the impact of this aspect of the material modeling on orientation sensitivity should be kept in mind when evaluating and parameterizing any other hardening model.

## 4. Summary and Conclusions

The systematic exploration of slight offsets in loading orientations within crystal plasticity finite element calculations of tensile deformation revealed a pattern of material response that is significant in the context of model parameterization. Minuscule offsets of ~0.1°, which unavoidably occur during experimental tests, significantly change the calculation results in terms of lattice rotation, slip system activation, and overall stress response of the material, especially for high symmetry orientations and high latent hardening ratios. The main conclusions are the following:

- With a latent hardening ratio greater than unity, misorientations as small as 0.1˚ resulted in significant deviations in the stress response of single crystal calculations for higher symmetry tensile orientations, including [001], [111], [101], and [102], but excluding [112] and [213].
- The misorientation effect on material stress response is linked to an increase of mean lattice reorientation and a decrease of the effective number of active slip systems during deformation relative to the direct loading case.
- The latent hardening ratio plays a critical role in determining slip system activity near multi-slip conditions, with values less than unity mitigating the misorientation effect that manifests strongly for the standard value of 1.4.
- Due to its potential for substantial impact on apparent stress-strain response, misorientation effects need to be considered when CPFE models are parameterized on multi-slip oriented single crystal data, whether new or extant in the literature, with the range of suitable offset magnitudes depending on the strength of slip system interactions in the chosen hardening model.

**Acknowledgments:**

This work was financially supported by the U. S. Department of Energy (DOE) via award no. DE-FE0031553. Computations were performed on The Pennsylvania State University's Institute for Computational and Data Sciences' Roar supercomputer. JDS acknowledges support from the Department of Energy National Nuclear Security Administration Stewardship Science Graduate Fellowship, provided under cooperative agreement number DE-NA0003960.

**Data availability statement:**

The raw and processed data required to reproduce these findings are available from https://doi.org/10.5281/zenodo.10684943.

# References

[1] F. Roters, P. Eisenlohr, L. Hantcherli, D.D. Tjahjanto, T.R. Bieler, D. Raabe, Overview of constitutive laws, kinematics, homogenization and multiscale methods in crystal plasticity finite-element modeling: Theory, experiments, applications, Acta Mater. 58 (2010) 1152–1211. https://doi.org/10.1016/j.actamat.2009.10.058.

[2] S. Nemat-Nasser, L. Ni, T. Okinaka, A constitutive model for fcc crystals with application to polycrystalline OFHC copper, Mech. Mater. 30 (1998) 325–341. https://doi.org/10.1016/S0167-6636(98)00055-6.

[3] S. El Shawish, L. Cizelj, Combining Single- and Poly-Crystalline Measurements for Identification of Crystal Plasticity Parameters: Application to Austenitic Stainless Steel, Crystals. 7 (2017) 181. https://doi.org/10.3390/cryst7060181.

[4] M.G. Lee, H. Lim, B.L. Adams, J.P. Hirth, R.H. Wagoner, A dislocation density-based single crystal constitutive equation, Int. J. Plast. 26 (2010) 925–938. https://doi.org/10.1016/j.ijplas.2009.11.004.

[5] R.S. Davis, R.L. Fleischer, J.D. Livingston, B. Chalmers, Effect of orientation on the plastic deformation of aluminum single crystals and bicrystals, JOM. 9 (1957) 136–140. https://doi.org/10.1007/BF03398471.

[6] R.J. Asaro, Crystal Plasticity, J. Appl. Mech. 50 (1983) 921. https://doi.org/10.1115/1.3167205.

[7] D. Peirce, R.J. Asaro, A. Needleman, An analysis of nonuniform and localized deformation in ductile single crystals, Acta Metall. 30 (1982) 1087–1119. https://doi.org/10.1016/0001-6160(82)90005-0.

[8] U.F. Kocks, The relation between polycrystal deformation and single-crystal deformation, Metall. Mater. Trans. B. 1 (1970) 1121–1143. https://doi.org/10.1007/BF02900224.

[9] A.M. Cuitino, M. Ortiz, Computational modelling of single crystals, Model. Simul. Mater. Sci. Eng. 1 (1993) 225–263. https://doi.org/10.1088/0965-0393/1/3/001.

[10] A. Arsenlis, D.M. Parks, Modeling the evolution of crystallographic dislocation density in crystal plasticity, J. Mech. Phys. Solids. 50 (2002) 1979–2009. https://doi.org/10.1016/S0022-5096(01)00134-X.

[11] B. Devincre, T. Hoc, L.P. Kubin, Collinear interactions of dislocations and slip systems, Mater. Sci. Eng. A. 400–401 (2005) 182–185. https://doi.org/10.1016/j.msea.2005.02.071.

[12] J.L. Dequiedt, C. Denoual, R. Madec, Heterogeneous deformation in ductile FCC single crystals in biaxial stretching: the influence of slip system interactions, J. Mech. Phys. Solids. 83 (2015) 301–318. https://doi.org/10.1016/j.jmps.2015.05.020.

[13] R. Madec, The Role of Collinear Interaction in Dislocation-Induced Hardening, Science. 301 (2003) 1879–1882. https://doi.org/10.1126/science.1085477.

[14] A. Ma, F. Roters, A constitutive model for fcc single crystals based on dislocation densities and its application to uniaxial compression of aluminium single crystals, Acta Mater. 52 (2004) 3603–3612. https://doi.org/10.1016/j.actamat.2004.04.012.

[15] J. Kuhn, J. Spitz, P. Sonnweber-Ribic, M. Schneider, T. Böhlke, Identifying material parameters in crystal plasticity by Bayesian optimization, Optim. Eng. 23 (2022) 1489–1523. https://doi.org/10.1007/s11081-021-09663-7.

[16] P.G. Heighway, J.S. Wark, Slip competition and rotation suppression in tantalum and copper during dynamic uniaxial compression, Phys. Rev. Mater. 6 (2022) 043605. https://doi.org/10.1103/PhysRevMaterials.6.043605.

[17] L.A. Zepeda-Ruiz, A. Stukowski, T. Oppelstrup, N. Bertin, N.R. Barton, R. Freitas, V.V. Bulatov, Atomistic insights into metal hardening, Nat. Mater. 20 (2021) 315–320. https://doi.org/10.1038/s41563-020-00815-1.

[18] T. Nguyen, S.J. Fensin, D.J. Luscher, Dynamic crystal plasticity modeling of single crystal tantalum and validation using Taylor cylinder impact tests, Int. J. Plast. 139 (2021) 102940. https://doi.org/10.1016/j.ijplas.2021.102940.

[19] Y. Huang, A user-material subroutine incorporating single crystal plasticity in the ABAQUS finite element program, Harvard, 1991.

[20] T. Takeuchi, Work Hardening of Copper Single Crystals with Multiple Glide Orientations, Trans. Jpn. Inst. Met. 16 (1975) 629–640. https://doi.org/10.2320/matertrans1960.16.629.

[21] R. Madec, L.P. Kubin, Dislocation strengthening in FCC metals and in BCC metals at high temperatures, Acta Mater. 126 (2017) 166–173. https://doi.org/10.1016/j.actamat.2016.12.040.

[22] J.D. Shimanek, S.-L. Shang, A.M. Beese, Z.-K. Liu, Insight into ideal shear strength of Ni-based dilute alloys using first-principles calculations and correlational analysis, Comput. Mater. Sci. 212 (2022) 111564. https://doi.org/10.1016/j.commatsci.2022.111564.

[23] J.L. Bassani, T.-Y. Wu, Latent Hardening in Single Crystals II. Analytical Characterization and Predictions, Proc. R. Soc. Math. Phys. Eng. Sci. 435 (1991) 21–41. https://doi.org/10.1098/rspa.1991.0128.

**Figures**

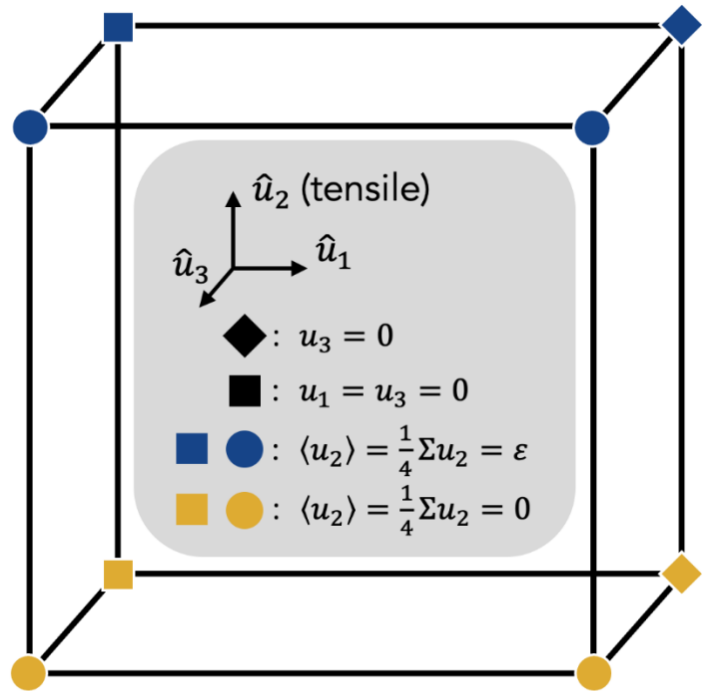

Figure 1. Schematic representation of the single element finite element model with boundary conditions in terms of displacements $u_i$ suitable for single crystal tensile strain $\varepsilon$ along $\hat{u}_2$.

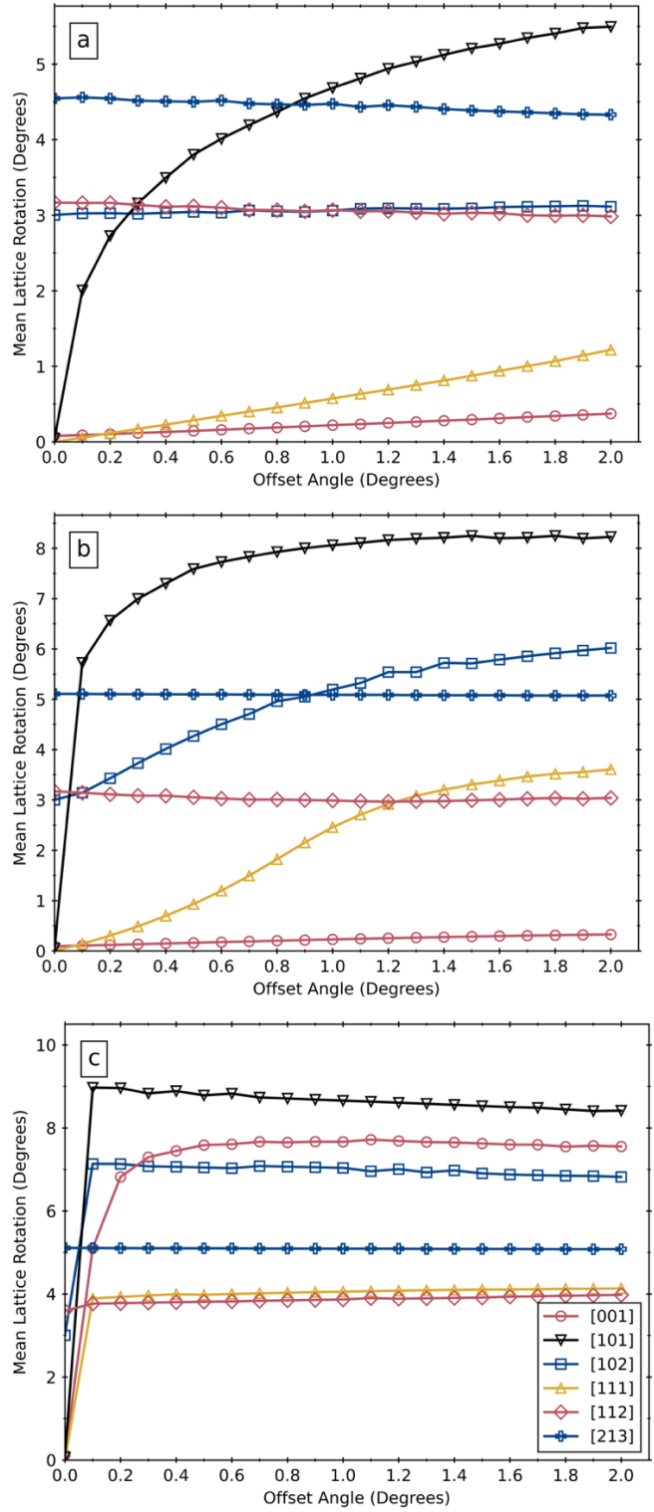

Figure 2. Magnitude of overall lattice rotation with respect to offset for each orientation for *q*-values of 0.8 (a), 1.0 (b), and 1.4 (c).

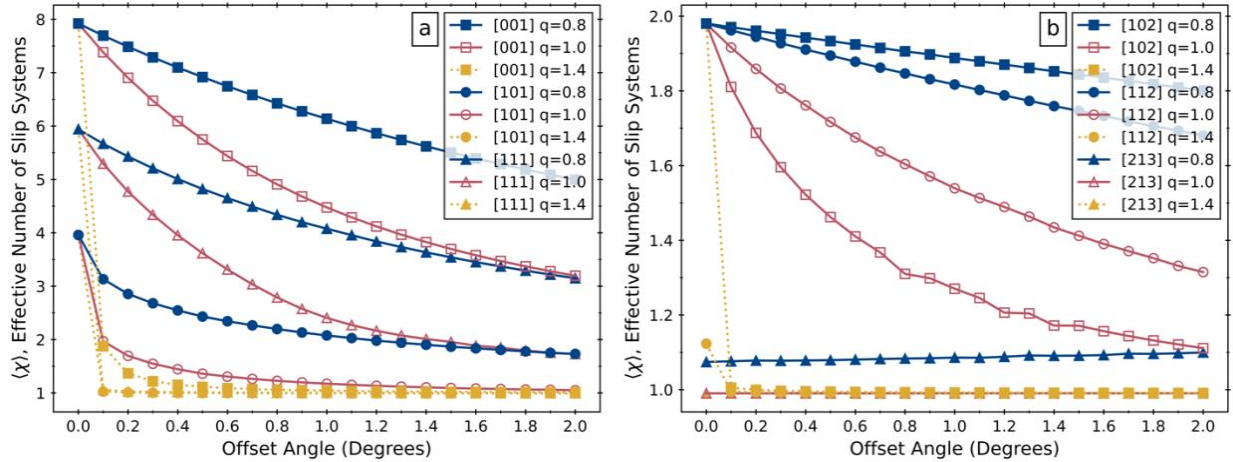

Figure 3. Mean effective number of active slip systems with respect to offset angle for all orientations, split into high-symmetry (a) and low-symmetry (b) loading orientations. Note that these plots are split by symmetry instead of by $q$-value and therefore use a different symbol mapping than Figure 1.

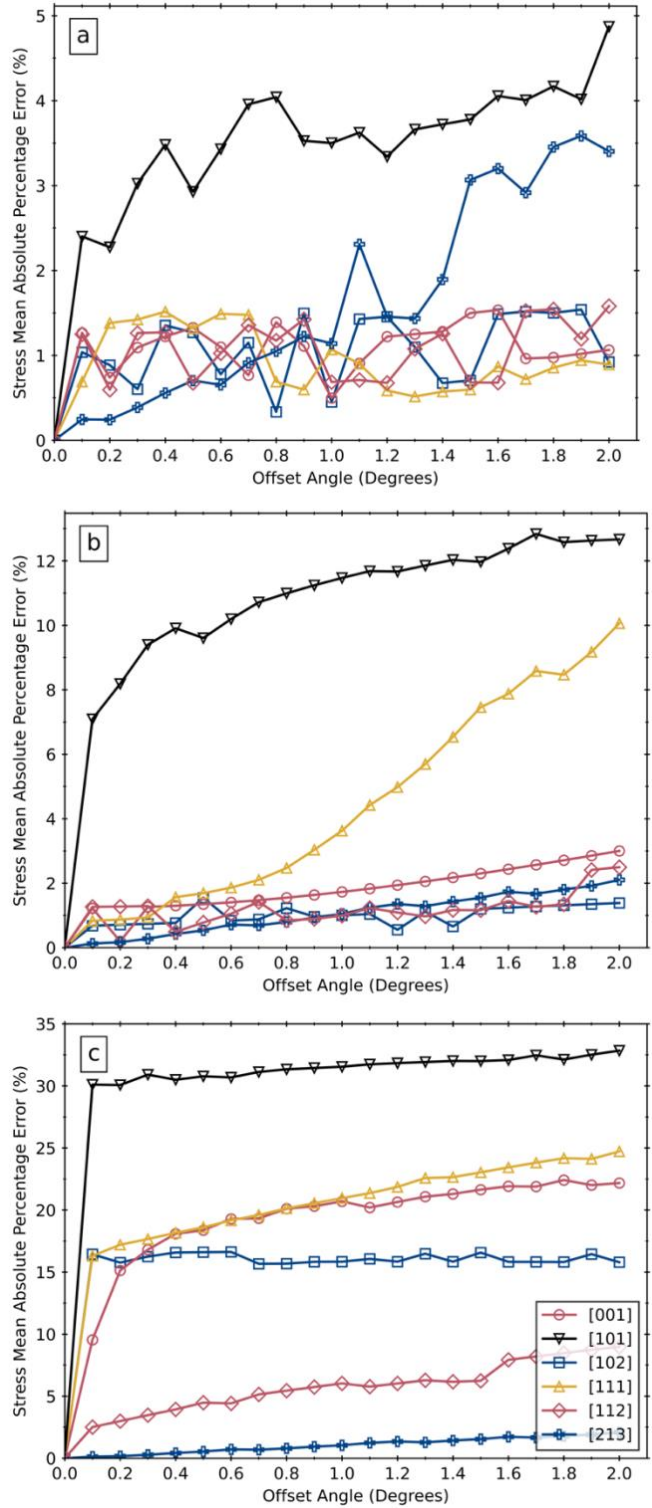

Figure 4. Relative error with respect to offset angle for all considered initial orientations and $q$-values of 0.8 (a), 1.0 (b), and 1.4 (c).

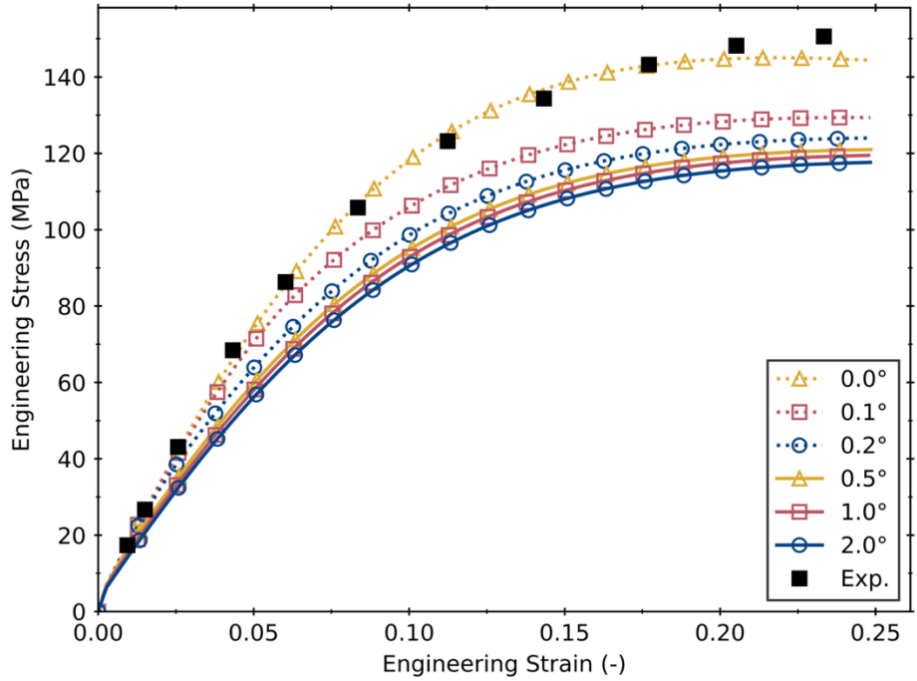

Figure 5. Engineering stress-strain curves showing the effect of several assumed misorientation magnitudes from the nominal [001] loading. Calculations used a latent hardening ratio of 1.4 and were calibrated based on the 0° offset case to tensile deformation data on room temperature Cu from Takeuchi [20], which are shown as black squares.

**Tables**

Table 1. Hardening parameters fit to single crystal Cu data from ref. [20] using $q = 1.4$ and no offset of the [001] loading axis.

| Parameter | Value | Units |
|---|---|---|
| $C_{11}$ | 159 | GPa |
| $C_{12}$ | 129 | GPa |
| $C_{44}$ | 29 | GPa |
| $\dot{\gamma}_0$ | 0.001 | – |
| $n$ | 50 | – |
| $\tau_0$ | 1 | MPa |
| $h_0$ | 192 | MPa |
| $\tau_s$ | 52 | MPa |

Supplemental Material for:

Effects of Misorientation on Single Crystal Plasticity by Finite Element Methods

John D. Shimanek[1,*], Zi-Kui Liu[1], Allison M. Beese[1,2,*]

[1] Department of Materials Science and Engineering, The Pennsylvania State University, University Park, PA, 16802, USA

[2] Department of Mechanical Engineering, The Pennsylvania State University, University Park, PA 16802, USA

* Corresponding authors: jshimanek@psu.edu, amb961@psu.edu

Model Geometry

The single element model with boundary conditions described in the main text was compared to a more explicit geometry constructed for wire tension simulations. The model geometry consisted of a long cylinder meshed into 145 elements per layer for 142 layers along the tensile direction. The same Cu material parameters were used, and the lattice orientation information was extracted from the central element of this larger "full geometry" model. Unlike the single element model, the top and bottom faces of the full geometry model are constrained to be flat, perpendicular to the applied loading. For loading along [101] with and without a 0.1° offset, the lattice reorientation magnitude is shown in Figure S1 and the stress-strain behavior in Figure S2. Also included in both figures are comparable metrics from the single element geometry using the same material parameters, showing a close match in all cases.

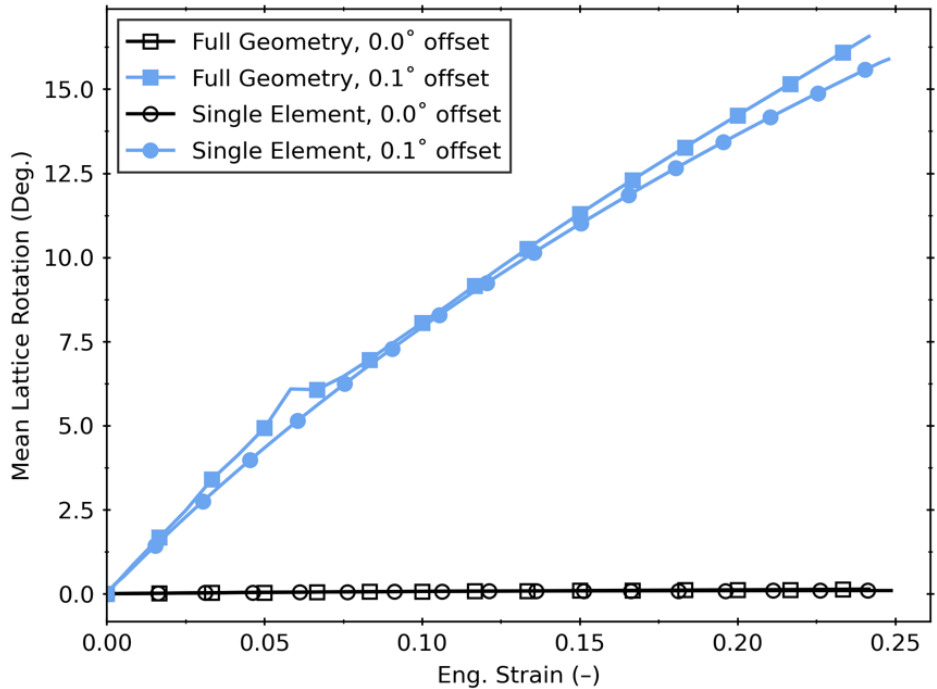

**Figure S1.** Lattice rotation as a function of axial tensile strain for the full geometry and single element models with and without a slight offset from [101] towards [213] loading. For the full geometry model, lattice orientation information was extracted from its center element.

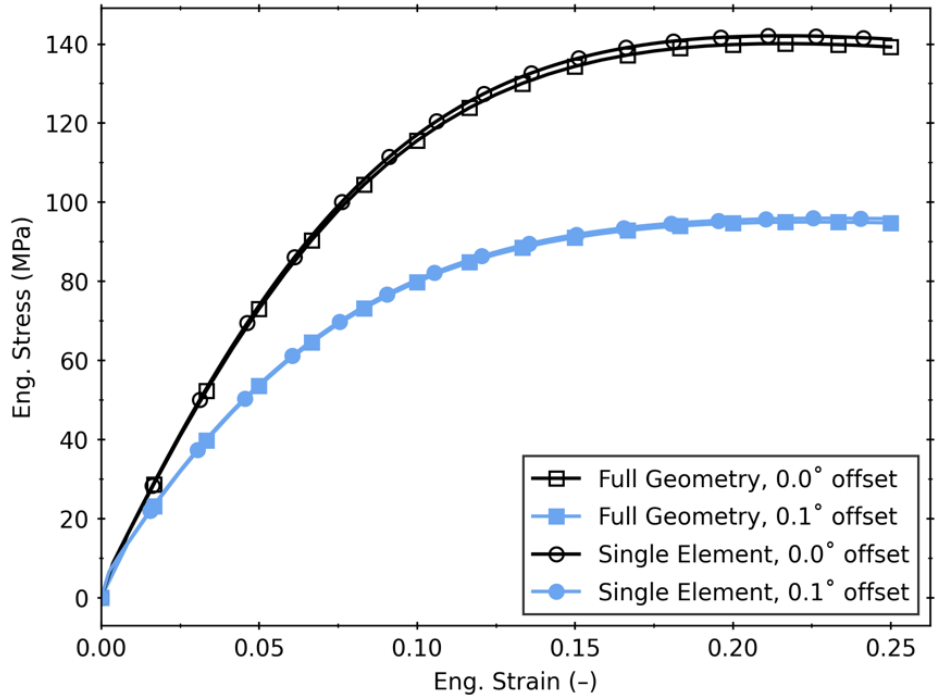

**Figure S2.** Engineering stress-strain comparison for the full geometry and single element models with and without a slight offset of [101] towards [213] loading.

Model Parameters

The model parameters used for the results presented in the main text are given in its Table 1. Those parameters were adjusted to use a latent hardening ratio of $q = 1.4$ while the remaining parameters were fit to experimental tensile data for a Cu single crystal oriented along [001], as performed by Takeuchi [1]. Additional parameterizations can be performed using more orientations from the same study, which is important for controlling against impurity level and experimental details that could influence the results between datasets. With the latent hardening ratio as an additional free parameter, the best fits to the orientations [001], [111], and [112] are shown below in Figure S3, with parameters given in Table S1.

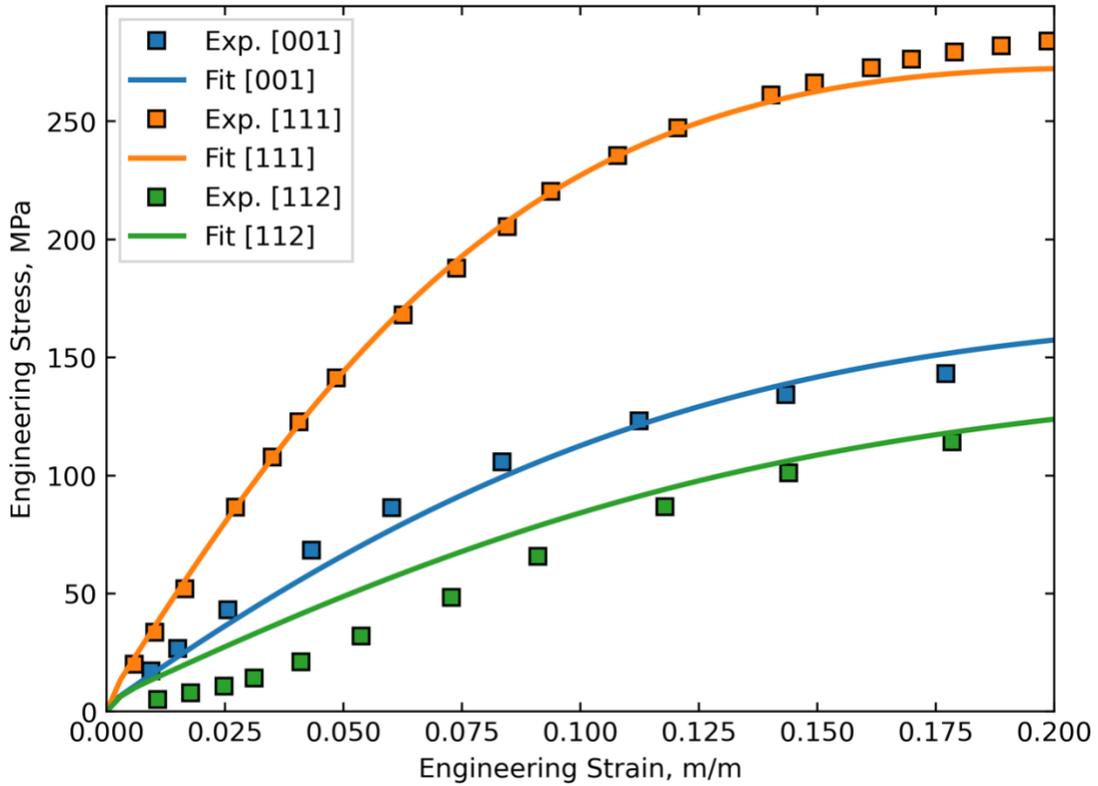

**Figure S3.** Engineering stress-strain curves used to find material parameters for the hardening model. Experimental data from Ref. [1].

**Table S1.** Model parameters for the stress-strain behavior shown in Figure S3.

| Parameter | Value |
|---|---|
| $h_0$ | 140. [MPa] |
| $\tau_s$ | 55.0 [MPa] |
| $q$ | 1.7 [–] |

Data from additional orientations can be added, with the relatively simple model showing a worse fit for all the data than for any individual orientation. Reasonable ordering is seen across most orientations. For multiple slip orientations, the hardening rate is over-predicted at the highest strains (see [111] and [112], e.g.). For the single slip orientation [213], the saturation stress is reasonably estimated while the initial hardening rate is over-predicted. However, the overall fit of the model to a wide range of data with a single parameter is reasonable while being simple enough to clearly demonstrate the misorientation effect that is the focus of the work.

Here, like for the three-orientation fitting, a slightly higher latent hardening ratio facilitated a better fit. Slight misorientations to [212] helped match its initial hardening rate and to [101] helped to match its saturation behavior. This suggests that consideration of possible misorientation effects, as outlined in the present manuscript, is necessary but not sufficient to perfectly deconvolute the extrinsic geometrical effects inherent to experimental from the intrinsic material deformation behavior.

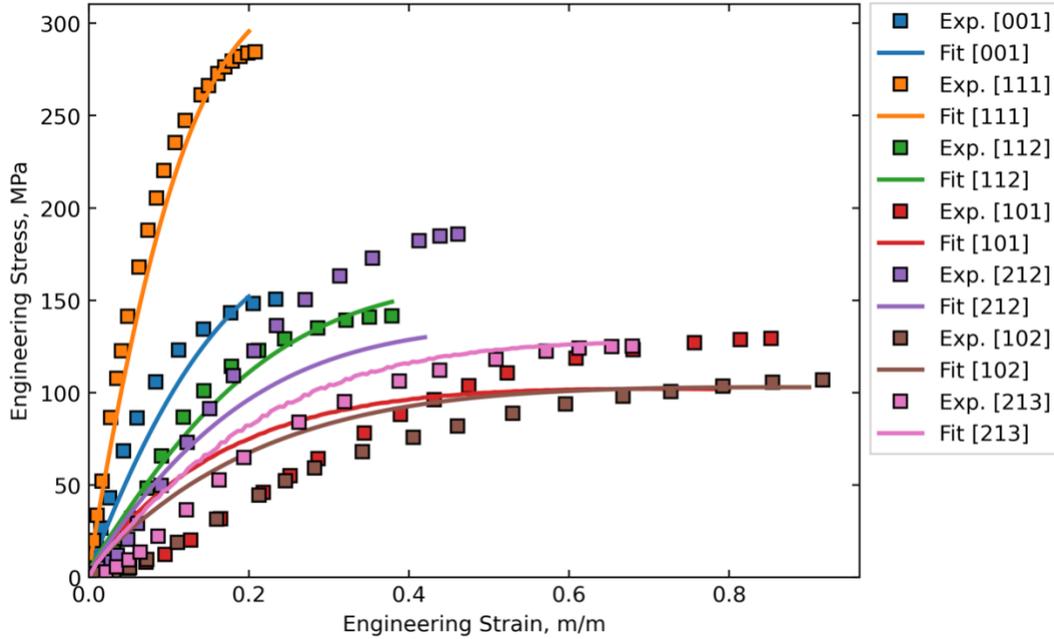

**Figure S4.** Engineering stress-strain curves for the current hardening model in comparison to the single crystal Cu data of Ref. [1].

**Table S2.** Model parameters for the curves shown in Figure S4.

| Parameter | Value |
|---|---|
| $h_0$ | 100. [MPa] |
| $\tau_s$ | 64. [MPa] |
| $q$ | 1.9 [–] |
| [212] offset | 1° |
| [101] offset | 1° |

The present fits can also be put in context of other available data, although some differences between the experimental datasets, due to purity levels and testing details, is to be expected. This is shown below in Figure S5, with some additional data on single crystal Cu for multiple samples of the same orientations dealt with in the present study [2]. Initial hardening rates show the most dispersion for [111] loading, although the present hardening model and the data it was parameterized for lie right in the middle of the group. For [011], the model does not capture the initial low hardening rate seen in either experimental dataset. For orientations like [001] and [012], the additional data imply a slightly closer alignment with this parameterized model, although the lower strain range prevents further comparison.

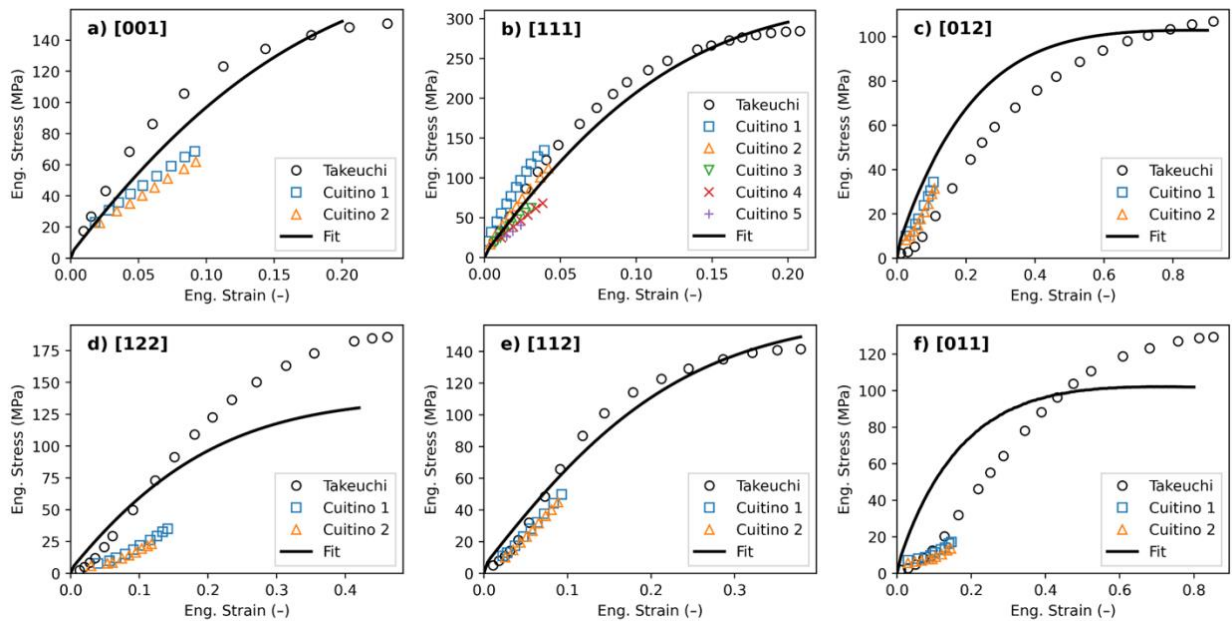

Figure S5. Engineering stress-strain curves using the crystal plasticity detailed in Table S2 ("Fit") but separated by nominal loading orientation and with the addition of some comparable Cu data ("Cuitino") [2,3].

References

[1] T. Takeuchi, Work Hardening of Copper Single Crystals with Multiple Glide Orientations, Trans. Jpn. Inst. Met. 16 (1975) 629–640. https://doi.org/10.2320/matertrans1960.16.629.
[2] A.M. Cuitino, M. Ortiz, Computational modelling of single crystals, Model. Simul. Mater. Sci. Eng. 1 (1993) 225–263. https://doi.org/10.1088/0965-0393/1/3/001.
[3] P. Franciosi, The concepts of latent hardening and strain hardening in metallic single crystals, Acta Metall. 33 (1985) 1601–1612. https://doi.org/10.1016/0001-6160(85)90154-3.